\documentclass[%
 aip,
amsmath,amssymb,
reprint,%
onecolumn,%
]{revtex4-1}

\newcommand{\be}{\vspace{-0.0802in}\begin{equation}}
\newcommand{\ee}{\vspace{-0.0702in}\end{equation}}

\usepackage{amscd}  

\usepackage{color}
\usepackage[margin=1.5in]{geometry}
\usepackage{graphicx}
\usepackage{dcolumn}
\usepackage{bm}
\usepackage{hyperref}
\usepackage{epstopdf}

\begin{document}


\title[]{Open or Closed?   Information Flow Decided \\by \\Transfer Operators \\ and \\ Forecastability Quality Metric}

\author{Erik M. Bollt}
\email{bolltem@clarkson.edu}
\affiliation{Department of Mathematics and Electrical and Computer Engineering and $C^3S^2$ 
the Clarkson Center for Complex Systems Science, Clarkson University, Potsdam, New York 13699, USA}

\date{\today}

\begin{abstract}

A basic systems question concerns the concept of closure, meaning autonomomy (closed) in the sense of describing the (sub)system as fully consistent within itself.  Alternatively, the system may be nonautonomous (open) meaning it receives influence from an outside coupling subsystem.  Information flow, and related causation inference, are tenant on this simple concept.  We take the perspective of Weiner-Granger causality, descriptive of a subsystem forecast quality dependence on considering  states of another subsystem.  Here we  develop a new direct analytic discussion, rather than a data oriented approach.  That is, we refer to the underlying Frobenius-Perron transfer operator that moderates evolution of densities of ensembles of orbits, and two alternative forms of the restricted Frobenius-Perron (FP) operator, interpreted as if either closed (determinstic FP) or not closed (the unaccounted outside influence seems stochastic and correspondingly the stochastic FP operator).   From this follows contrasting the kernels of the variants of the operators, as if densities in their own rights.  However, the corresponding differential entropy to compare by Kulback-Leibler divergence, as one would when leading to transfer entropy, becomes ill-defined.  Instead we build our Forecastability Quality Metric (FQM) upon the ``symmetrized" variant known as Jensen-Shanon divergence, and also we are able to point out several useful resulting  properties that result.  We illustrate the FQM by a simple coupled chaotic system.  For now, this analysis is a new theoretical direction, but we describe data oriented directions for the future.

\end{abstract}

\maketitle

{\bf Causation inference in the sense of G-causality (Granger causality) refers to the concept of reduction of variance.  That is to answer the basic question, does system X allow for sufficient information regarding  forecasts of future states of system X or is there improved forecasts with observations from system Y.  If the later is the case, then we declare that X is not closed, as it is receiving influence, or information, from system Y.  Such a reduction of uncertainty perspective of causal influence is not identical to the concept of allowing perturbations and experiments on two systems to decide what changes influence each other.  Methods such as Ganger causality, transfer entropy, causation entropy, and even cross correlation method, each are premised on the concept of alternative formulations of the forecasting question, with and without considering the influence of an external state. Thus the idea is to decide if a system is open or closed.  Here we assert that the underlying transfer operator, the Frobenius-Perron operator, that moderates not the evolution of single initial conditions, but the evolution of ensembles of initial conditions allows for a direct and sensible analysis of information flow to decide the question of open or closed.  Specifically a restricted form of the transfer operator to a subsystem, queried as if with or without the states of the other subsystem(s), allows for a new analytically tractable formulation of the question of information flow.  In this philosophy, the exterior system  becomes an ``unknown" influence onto the interior system, and as such it becomes formulated as a stochastic influence and correspondingly stochastic transfer operator.  In this philosophy it becomes clear that even though the exterior system may be deterministic, it appears stochastic within focus on the interior system.  As an explicit measurement for this concept we build a Forecastability Quality Metric (FQM) based on Jensen-Shannon divergence, applied directly to alternative forms of the transfer operator, noting that a transfer entropy like application of Kulback-Leibler divergence would be impossible.  However, this choice of metric like measurement allows for several especially elegant properties that we annunciate here.  Application is describe and future empirical directions are described.}


\section{Introduction}

A basic question in information flow is to contrast versions of reality for a dynamical system, whether a subcomponent is closed versus whether there is an outside influence by another component.  The details of how this question is posed and how it is decided gives rise to various versions of concepts of information flow, and related to causation inference, including for example the celebrated Nobel work \cite{hendry2004nobel} behind Granger Causality \cite{granger1969investigating, granger1988some, barnett2009granger} and closely related  by N.~Wiener, \cite{wiener1956modern}, as well as the popular transfer entropy, \cite{schreiber2000measuring, bollt2013applied} and related causation entropy to furthermore uncover the differences between direct and indirect influences, \cite{sun2014causation, sun2015causal, sun2014identifying, cafaro2015causation} , as well as direct forecast methods including cross correlation method \cite{sugihara2012detecting}.  Generally when discussing computational causation as derived from data  measured from a dynamical system, with some variant of the question as cited above, we shall interpret information flow  as a {\it measurement of reduction of uncertainty associated with forecasts} from a subcomponent when considering including a fuller model by associating another  subcomponent; when that other subcomponent allows for better forecasts it is decided that the subsystem is not closed, as it must be receiving information from the other subsystem.  Often information flow, as a concept of reduction of uncertainty was discussed  as related to concepts of causation.  Causation inference has overlapping philosophical roots \cite{russell1912notion}, and we have also allowed our own previous writings on these topics to overlap these distinct concepts,  but here in this paper we will simply discuss information flow as a form of reduction of uncertainty.  In fact there is a beautiful connection between Granger causality and transfer entropy in the special case of Gaussian noise \cite{barnett2009granger}.  Furthermore in \cite{barnett2009granger} is distinguished the concept of Wiener-Granger causality (G-causality) \cite{granger1969investigating,wiener1956modern} that between two inter-dependent variables, $X$ and $Y$, in a statistical sense  that $Y$- G-causes $X$ if measurements of $Y$ can improve forecast of future values of measurements of $X$ beyond that possible by measurements of $X$ alone; this is what we mean by reduction of uncertainty and this is the nonintervention philosophy that we will maintain here.  This is in contrast to the related but distinct  concept of ``physically instantiated causal relationship" \cite{barnett2009granger}  in a sense that can only be truly uncovered by perturbations to the system, as the statistics of causation by interventions and observations described in Pearl's extensive work enunciates \cite{pearl2009causality}.  

Most studies on information flow are in terms of data and the statistical inference concepts cited above, \cite{granger1969investigating, granger1988some, barnett2009granger} sometimes by information theoretic methods, \cite{schreiber2000measuring, bollt2013applied, sun2014causation, sun2015causal, sun2014identifying, cafaro2015causation}, although most notably see Liang-Kleeman, \cite{liang2005information} as a more analytical approach that involves both the dynamical system and the concept of information flow.  Also our own  prior work relates synchronization as a process of sharing information, \cite{bollt2012synchronization}.  This current work then builds on \cite{bollt2012synchronization} in that we refer directly to transfer operators when describing the degree to which a system may or may not be open.

In this paper we describe a new approach formalism of analysis of the underlying concept of reduction of uncertainty in terms of evolution of densities.
It is the Frobenius-Perron operator whereby the question of how ensembles (densities) of initial conditions evolve under orbits of a dynamical system, \cite{lasota1985probabilistic, bollt2013applied}.  Within this  framework of transfer operators, we may recast the question of information flow by more rigorously presenting the two versions of the basic question, which is to decide one of the two  alternatives:
\begin{itemize}
\item Is the subsystem closed? \\
\item Does the subsystem receive influence from another subcomponent?
\end{itemize}

Our own previous work considered the relationship of evolution of densities as moderated by the Frobenius-Perron operator, together with the information theoretic question of information flow by transfer entropy \cite{bollt2012synchronization, bollt2013applied}.  However the details of our previous work was discussed in terms of estimating the associated probability density functions at steady state, and furthermore, through estimation of the transfer operator's action on the space of densities by the famous Ulam-Galerkin's methods of projection on to a linear subspace, $\Delta_N$, as  $P:L^2(\Omega)\rightarrow \Delta_N\subset L^2(\Omega)$ to describe finite matrix computations.  There is a long history to the Ulam's method, \cite{ulam1960problems,li1976finite, froyland1999ulam, hunt1998unique, murray2010ulam, matousek2008using, bollt2013applied, lasota1985probabilistic, billings2001probability, bollt2012measurable,  bollt2000controlling,  bollt2002manifold}, but this approach generally relies on covering the space with boxes, and estimating probabilities in a histogram-like fashion at steady state so that the estimations can be statistically stationary.  This current work takes a signficant departure of the theme of steady state and we do so within the scope of transfer operators directly with a new interpretation of the external influences that can be exactly and analytically described as a random variable like term in a variation of the standard Frobenius-Perron operator.

A unique outcome of our study is that a more standard transfer entropy approach by a Kullback-Leibler, $D_{KL}$,  interpreted in terms of the kernel of the transfer  operator, is unbounded.    Instead use the so-called symmetrized version of KL-divergence, called the Jensen-Shannon divergence, $D_{JS}$. Not only does this approach fix an otherwise unpleasant nonconvergence problem, but also it brings with it several beautiful new properties and interpretations that underly the theory special to the JS divergence.  With these new interpretations in mind, we call this variant of information flow, the Forecastability Quality Metric, written $FQM_{y\rightarrow x}$ in analogy to the notation one uses typically for transfer entropy, $T_{y\rightarrow x}$ between subsystem $y$ and subsystem $x$.  

The work presented herein could be considered theoretical in nature, marrying the theories of transfer operators, statistics and information theory in a unique way to well define a concept of information flow in terms defining the difference between closed and not closed subsystems.  Thus in standing up a new perspective on what these questions mean interpreting within the formal language of these disparate fields, we hope we can better sharpen the general understanding of these important questions.  Nonetheless we will point out at the end of the paper directions in which this perspective could be turned into a data oriented methodology.  
 
\section{Basic Problem Setup}

A most basic version of the discussion of a full system with subcomponents follows consideration of  two coupled oscillators,
\begin{eqnarray}\label{identa}
x_{n+1}&=&f_1(x_n)+\epsilon_1 k(x_n,y_n) \nonumber  \\
y_{n+1}&=&f_2(y_n)+\epsilon_2 k(y_n,x_n).
\end{eqnarray}
We might ask if the ``x-oscillator" is ``talking to" the ``y-oscillator", and vice-versa.  Defining the concept of ``talking to" may be made in various forms.   Avoiding philosophical notions,   we take the perspective of predictability, by asking if $x$ variables improve forecasts of future states of $y$-variables better than considering just $y$ variables alone, in the sense of reduction of uncertainty, thus G-causality.


Now we recast the typical symmetrically coupled problem, Eq.~(\ref{ident}), to a general form
of a partitioned dynamical systems on a skew product space $X\times Y$,
\begin{equation}\label{skewproductstates}
T:\Omega_X\times \Omega_Y\rightarrow \Omega_X\times \Omega_Y.
\end{equation}
This emphasizes that the full system is a single dynamical system where the phase space is a skew product space, so  examples such as Eq.~(\ref{identa}) discuss information flow between the $\Omega_X$ and $\Omega_Y$ states. 
In this notation then, the two component coupled dynamical systems of the $x$ and $y$  component variables may be written,
\begin{equation}\label{form1}
T(x,y)=(T_x(x,y),T_y(x,y)).
\end{equation}
where,
\begin{eqnarray}\label{form2}
T_x:X\times Y &\rightarrow& X \nonumber \\
	x_n &\mapsto& x_{n+1}=T_x(x_n,y_n), \nonumber \\
T_y:X\times Y &\rightarrow& Y \nonumber \\
	y_n &\mapsto&y_{n+1}=T_y(x_n,y_n).
\end{eqnarray}
In the case of Eq.~(\ref{identa}), let,
\begin{eqnarray}\label{form3}
T_x(x_n,y_n)&=&f_1(x_n)+\epsilon_1 k(x_n,y_n), \mbox{ and} \nonumber \\ T_y(x_n,y_n)&=&f_2(y_n)+\epsilon_2 k(y_n,xn).
\end{eqnarray}
The notation $x\in \Omega_X$ and $y\in \Omega_Y$ allows that each may be vector valued and generally  even  differing dimensionality.    We write, $\Omega=\Omega_X\times \Omega_Y$, but sometimes in the subsequent we will write $\Omega$ as the phase space of an unspecified transformation, and these phase spaces will also serve conveniently as outcome spaces when describing the dynamical systems as stochastic processes.

\section{Information Flow as Alternative Versions of Forecasts in Probability}

Information flow is premised on a simple question of comparing alternative versions of forecasts, stated probabilistically.
We ask the question as to if two different probability distributions are the same, or different,  which can be stated \cite{bollt2013applied}
\begin{equation}\label{dev}
P(x_{n+1}|x_n)= \hspace{-0.08in}? \hspace{0.08in} P(x_{n+1}|x_n,y_n),
\end{equation}
and if they are different, the degree to which they are different.  This describes a degree of deviation from a Markov-property.

\subsection{Information Flow as Transfer Entropy}

Specifically the transfer entropy \cite{schreiber2000measuring}  measures  deviation from the Markov-property question, Eq.~(\ref{dev}) using the Kullback-Leibler divergence,
\begin{equation}\label{dev2}
T_{y\rightarrow x} =D_{KL}(p(x_{n+1}|x_n)||p(x_{n+1}|x_n,y_n)),
\end{equation}
in terms of the probability distributions associated with the probabilities of Eq.~(\ref{dev}).  A useful outcome in using this entropy-based measure to describe deviation from Markov-ness, is that the answer is naturally describing information flow in units of bit/second.  In subsequent sections we will point out problems in Kullback-Leibler divergence that are solved by answering the same question with the Shannon-Jensen divergence instead, with some lovely special properties to also follow.  Generally the transfer entropy was defined \cite{schreiber2000measuring} in terms of $k$-previous states in each variable, but we take this, simplification to single prior states to be associated with the related problem of true embedding in delay variables, \cite{eckmann1985ergodic, sauer1991embedology, bollt2000model}.

Note that the probability density functions written in Eqs.~(\ref{dev})-(\ref{dev2}) respectively are not generally the same functions and as noted by the differing arguments.  Furthermore, they need not be assumed to be steady state probabilities; this is an important distinction in the course of this paper as departure from many previous works in information flow.  Instead generally consider them as nonequilibrium functions representing the state of probabilities of ensembles of orbit states $(x_n,y_n)$, at time-$n$, following a random ensemble of initial states, $(x_0,y_0)$, but observed at time $n$.

Here we will keep with the description that the outcome spaces may be continuous and state the differential entropy version of a Kullback-Leibler divergence definition for transfer entropy.
A general definition that suits our purposes is as follows.  Let outcome space $\Omega$ have a measure $\mu$, so that probability measures $P_1$ and $P_2$ are absolutely continuous to $\mu$, so that $p_1=\frac{d P_1}{d \mu}$ and $p_2=\frac{d P_2}{d \mu}$, then $D_{KL}(P_1||P_2)$ may be written,
\begin{equation}
D_{KL}(P_1||P_2)=\int_\Omega p_1 \log \frac{p_1}{p_2} d\mu=-H(p_1)- \int_\Omega p_1 \log {p_2} d\mu,
\end{equation}
but we will allow the abuse of notation to write the KL-divergence in terms of the pdf's as the arguments, $D_{KL}(p_1||p_2)$.
Therefore, when there are continuous state spaces, let, 
\begin{eqnarray}
T_{y\rightarrow x}&=&D_{KL}(p(x_{n+1}|x_n,y_n)||p(x_{n+1}|x_n)= \nonumber \\
&=&
\int_\Omega p(x_{n+1}|x_n,y_n)\log [{p(x_{n+1}|x_n,y_n)}-{p(x_{n+1}|x_n)}] d\Omega, 
\end{eqnarray}
and in this integral, $\Omega=X_n \times X_{n+1}\times Y_n$.  The expression for $T_{x\rightarrow y}$ is similar ,
\begin{equation}
T_{x\rightarrow y}=D_{KL}(p(y_{n+1}|x_n,y_n)||p(y_{n+1}|y_n).
\end{equation}

There is however a significant technical difficulty with using the Kullback-Leibler divergence in this way as generally $D(p_1||p_2)$ is only bounded if the support of $p_1$ is contained in the support of $p_2$, \cite{cover2012elements}.  This turns out to be untrue in a natural interpretation of transfer entropy that follows when directly approaching the description of the densities by the kernel of the transfer operators.  This will motivate out fix to the problem by developing the $FQM_{y\rightarrow x}$.  Also the usual interpretation, to assign $0 \log \frac{0}{0}=0$ is useful here to emphasize continuity.

\section{Evolution of Densities by of Ensembles of Initial Conditions by the Frobenius-Perron operator}

The evolution of single initial conditions proceeds by the mapping  $T:\Omega_X\times \Omega_Y\rightarrow \Omega_X\times \Omega_Y$. But the evolution of many initial conditions all at once follows evolution of ensemble densities of many states both before and after the mapping is applied. The Frobenius-Perron operator is defined to describe the associated dynamical system for evolution of densities.   First we review this theory, and then we will specialize the concepts to both the full problem and the marginalized problem, both considering with and without the coupling term.  What is especially new here, is that in the coupled case, the coupling influence of the other dynamical system enters in a way that may be interpreted as a stochastic perturbation, so associated to the stochastically perturbed Frobenius-Perron operator.
\subsection{The Deterministic Frobenius-Perron Operators}

Remarkably, while even for a nonlinear dynamical system 
\begin{equation}\label{DS}
F:M  \rightarrow  M, 
\end{equation}
 which may even be chaotic, the one-step action of the map in the space of (ensembles of initial conditions) densities is that of a linear transfer operator \cite{lasota1985probabilistic,bollt2013applied}, for a phase space, $M\subset {\mathbb R}^n$.  The Frobenius-Perron operator generates an associated linear dynamical system on the space of densities.
\begin{equation}\label{FP}
P_F: L^1(M)\rightarrow L^1(M)
\end{equation}
defined by,
\begin{equation}\label{FP1D}
P_F[\rho](z)=\int_M \delta(z-F(s)) \rho(s) ds= \sum_{\{s:F(s)=z\}} \frac{\rho(s)}{|F'(s)|}
\end{equation}
where the sum is taken over all pre-images, $s$, when the map has a multiply branched ``inverse".  Note also that in the case of a multi-variate transformation $F$, $m>1$, then the term $\sum_{\{s:F(s)=z\}} \frac{\rho(s)}{|F'(s)|}$ is instead replaced by $\sum_{\{s:F(s)=z\}} {\rho(s)}{|DF^{-1}(s)|}$, meaning the determinant of the Jacobian derivative matrix of the inverse of the map must be used.
While this infinite dimensional operator is typically not realizable in closed form, except for special cases \cite{lasota1985probabilistic,bollt2013applied}, there are matrix-methods in terms of approximating the action of the transformation as a stochastic matrix, and weak convergence to the true invariant density is called Ulam's method, \cite{Ulam, li1976finite,Froyland1,Froyland2,Boyarsky-Lou}, as a  technique to project this operator to a finite dimensional linear subspace  $\Delta_N\subset L^2(M)$ generated by the set characteristic functions supported over the partitioning grid \cite{li1976finite}.  The idea is that refining the grid yields weak approximants of invariant density.    The projection is exact when the map is ``Markov" using basis functions supported on the Markov partition, \cite{boyarsky2012laws, bollt2005markov, bollt2008basis}. Roughly speaking, the infinitesimal transfer operator, \cite{cvitanovic1991periodic},  $\cal{L}$$(s,z)=\delta(s-F(z))$, when integrated over a grid square $B_i$ which are small enough so that $DF(z)$ is approximately constant, then this action is approximated by a constant matrix entry $S_{i,j}$.    Under special assumptions on $F$, statements concerning quality of the approximation can be made rigorous.
Recently, many researchers have been using Ulam's method to describe global statistics of a dynamical system, \cite{dellnitz1997exploring, hunt1992approximation,Froyland1,Froyland2,Boyarsky-Lou} such as invariant measure, Lyapunov ezponent, dimension, etc.   A point of this paper is to get away from three major aspects of this kind of computation which are 1) the estimations based on the finite rank matrix computations, 2) the statistical approximations based on estimation of the matrix entries, 3) the inherently steady state  stationarity concept assumptions for collecting the statistics of $S_{i,j}$; those assumptions were previously built  into our own Ulam-Galerkin based approach to transfer entropy by Frobenius-Perron operator methods, \cite{bollt2012synchronization}.  Instead our descriptions will be in terms of the full integral describing the transfer operator adapted to notions of information flow, with no underlying assumption of steady state.

\subsection{The Stochastic Frobenius-Perron Operators}
Now consider the stochastically perturbed dynamical system,
\begin{eqnarray}\label{smap}
F_g:M & \rightarrow&  M,  \\
 \mbox{ } z &\mapsto &F(z)+\xi,  
\end{eqnarray}
where $\xi$ is a random variable with pdf $g$,
 which is applied
once per each iteration. The usual assumption at this stage is that the realizations  $\xi_n$ of $\xi$
added to subsequent iterations form an i.i.d. (identical independently distributed)
sequence, but since we are allowing for just one application of the dynamic process, the assumption is not necessary, and $g$ maybe simply be the distribution of $\xi_n$  at time $n$.
  If  $\xi$ is relatively small to $x$ then the
deterministic part $F$ has primary influence, but this is not even a necessary assumption for this stochastic Frobenius-Perron operator formalism.  Neither is a standard assumption for many stochastic analysis that require certain forms of the noise term, such as Gaussian distributed, as we do not require anything other than $g$ is a measurable function, which  likely is the  weakest kind of assumption possible.
The ``stochastic Frobenius-Perron operator'' has a similar form to the
deterministic case \cite{lasota1985probabilistic, bollt2013applied},
\begin{equation}
\label{FPnoise2}
P_{F_g}[\rho](z)=\int_M g(z-F(s)) \rho(s) ds,
\end{equation}
It is interesting to compare this integral kernel to the delta function in Eq.~(\ref{FP1D}). Now
 a stochastic kernel describes the pdf of the noise perturbation.  
We denote the stochastic Frobenius-Perron operator to be $P_{F_g}$, vice $P_{F}$ for no noise version in Eq.~(\ref{FP1D}).
In the case that the random map Eq.~(\ref{smap}) arises from the usual
continuous Langevin process, the infinitesimal generator of the
Frobenius-Perron operator for Gaussian $g$ corresponds to a
general solution
of a standard Fokker-Planck equation, \cite{lasota1985probabilistic}.  

Within the same formalism, we can also study multiplicative noise,
\begin{equation}
z\rightarrow \eta \tau(x),
\end{equation}
 modeling parametric noise).   It can be proved, \cite{santitissadeekorn2007infinitesimal, bollt2013applied} that the kernel-type
integral transfer operator is, 
\begin{equation}
{\cal K}(z,s)=g(z/F(s))/F(s).
\end{equation}
More generally,
the theory of random dynamical systems \cite{arnold2013random}  classifies those
random systems which give rise to explicit transfer operators with
corresponding infinitesimal generators, and there are well defined connections
between the theories of random dynamical systems and of stochastic
differential equations.  

\section{Interpreting Closure by Evolution of Density in Terms of Transfer Operators, with Associated Conditional Probability Densities}

Consider now the Frobenius-Perron operator Eq.~(\ref{FP1D}) term by term as associated with relevant  conditional and joint probabilities.  First let, $z=x-F(y)$, which upon substitution into Eq.~(\ref{FP1D}), yields, 
\begin{equation}\label{twv1}
\overline{\rho}(x)=P_F[\rho](x)=\int_M \delta(x-F(s)) \rho(s) ds= \int_M \frac{\delta(z)}{F'(F^{-1}(x-s))} \rho(F^{-1}(x-s)) ds.
\end{equation}
By a similar computation, with the same substitution, the stochastic version of the Frobenius-Perron operator, Eq.~(\ref{FPnoise2}) can be written, 
\begin{equation}\label{twv2}
\overline{\rho}(x)=P_{F_g}[\rho](x)=\int_M g(x-F(s)) \rho(s) ds=\int_M \frac{g(z)}{F'(F^{-1}(x-s))} \rho(F^{-1}(x-s)) ds.
\end{equation}
(Again if the transformation is multivariate, then the determinant of the Jacobian, or so-called Wronskian, must be used, ${|DF^{-1}(y)|}$).
Now we have written the new distribution of points as, $\overline{\rho}(x)$, evaluated at a point $x\in M$.  Notice that these Eqs.~(\ref{twv1})-(\ref{twv2}) are essentially the same in the special case that the distribution $g$ is taken to be a delta function, as if the noise has limited to zero variance in the sense of weak convergence.

Let us interpret these pdf functions as describing probabilities as follows.  It is useful at time $n$ to associate,
\begin{equation}\label{prob1}
P(X_n\in(x,x+dx))=\rho(x)dx,
\end{equation}
and
\begin{equation}\label{prob2}
P(X_{n+1}\in(x,x+dx'))=\overline{\rho}(x)dx',
\end{equation}
and $(x,x+dx')$ may denote small measurable sets containing at $x$ in the general multivariate scenario.

Take $\rho$ to be the probability distribution associated with samples of the ensemble of points along orbits, at time, $n$ and likewise $\overline{\rho}$, at time $n+1$.   
Interpreted in this way as a stochastic system (where the randomness is associated with the initial selection from the ensemble), depends on which version of the dynamics (with our without randomness, upon iteration) whether version, Eq.~(\ref{DS}) or Eq.~(\ref{smap}).  



Recall that 
 since, (by general conditional probability formula, $P(A|B)\cdot P(B)=P(A,B)$), or a chain statement for compound events, 
 \begin{equation}\label{int1}
 P(A,B,C)=P(A|B,C)\cdot P(B|C)\cdot P(C).
 \end{equation}
 Then let events be defined, 
 \begin{eqnarray}\label{int2}
 A&=&[X_{n+1}=x] \nonumber \\
 B&=&[X_n=F^{-1}(x-y)] \nonumber \\
 C&=&[Y_n=y]
 \end{eqnarray}
 For convenience we will now drop the formal descriptions of small intervals as $dx, dx', dy$ and the careful notation of probability events in intervals, as noted in Eqs.~(\ref{prob1})-(\ref{prob2}).  So more loosely in notation now, we describe,
\begin{eqnarray}\label{int3}
P(X_{n+1}=x|X_n=F^{-1}(x-y),Y_n=y) \cdot & P(X_n=F^{-1}(x-y)|Y_n=y) \cdot  P(Y_n=y)=   \nonumber \\
=\frac{1}{F'(F^{-1}(x-y))} \cdot  \rho(F^{-1}(x-y) \cdot  g(y), 
\nonumber \\
\end{eqnarray}
with the interpretation, 
\begin{eqnarray}\label{int4}
P(X_{n+1}=x|X_n=F^{-1}(x-y),Y_n=y)&=& \frac{1}{F'(F^{-1}(x-y))}  \\
P(X_n=F^{-1}(x-y)|Y_n=y)&=& \rho(F^{-1}(x-y)  \\
P(Y_n=y)&=& g(y). \label{int4b}
\end{eqnarray}
This interpretation allows us to compute a conditional entropy of evolution both with and without full consideration of externals to the partitioned subsystem effects.

\begin{figure}[htbp]
\centering
\includegraphics*[width=0.45\textwidth]{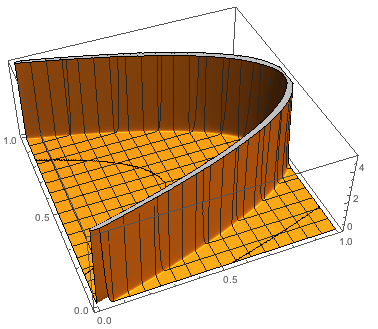}
\includegraphics*[width=0.45\textwidth]{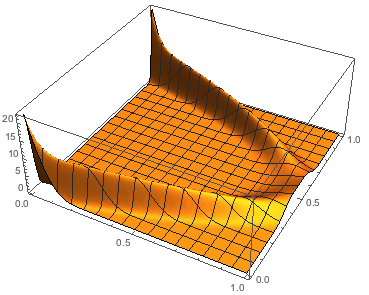}
\caption{Contrasting the kernel functions, (Left) $\delta(x-F(s))$ and (Right) $g(x-F(s))$, for contrasting versions of the transfer operator corresponding to a logistic map, Eqs.~(\ref{logistic1})-(\ref{logistic3}), for closed versus open versions of the concept of the primary question of the associated conditional probabilities question in Eqs.~(\ref{dev2})-(\ref{dev3}) as decided by the $FQM_{y\rightarrow x}$.
}\label{fig:1}
\end{figure}

To explicitly interpret a transfer entropy described seamlessly together with the evolution of densities derived by Frobenius-Perron transfer operator, we may be interested to understand the propensity of the mapping $F$ to move densities and then in this context we may therefore assume a specific simple form, $\rho$ is uniform.  Therefore in this context, recombining Eqs.~(\ref{int4}), (\ref{int4b}) suggest an interpretation,
\begin{eqnarray}\label{int5}
P(X_{n+1}=x,Y_n=y|X_n=F^{-1}(x-y))&=& q(x-F(y)),
\end{eqnarray}
with,
\begin{equation}\label{normalize}
q(x-F(y))=\frac{g(x-F(y))}{\int g(x-F(y)) dy},
\end{equation}
the integral kernel normalized as a probability distribution, for each $x$,

While this is not the same as the original question leading to transfer entropy, Eq.~(\ref{dev}),
$P(x_{n+1}|x_n)= \hspace{-0.08in}? \hspace{0.08in} P(x_{n+1}|x_n,y_n),$ we find comparing the kernel's corresponding to a system that is closed unto itself, versus that of a system that is receiving the information at each step by the action of the associated transfer operator, to be extremely informative.  But now as we see, this amounts to a slightly different but perhaps related question,
\begin{equation}\label{dev2}
P(x_{n+1}|x_n)= \hspace{-0.08in}? \hspace{0.08in} P(x_{n+1},y_n|x_n).
\end{equation}
These two alternative stories, closed, or open, of what may moderate the $x$-subsystem of the dynamical system,
\begin{equation}\label{dev3}
q(x-F(y))= \hspace{-0.08in}? \hspace{0.08in} \delta(x-F(y)),
\end{equation}
 distinguish the cases whether the $x$-subsystem is  closed, or if it is open - receiving information from the $y$-subsystem.  Therefore in the subsequent we will describe how to compare these, within the language of information theory.  See contrasting versions of Eq.~(\ref{dev3}) in Fig.~\ref{fig:1}, described in details as the example in Sec.~\ref{examplesec}.

\section{Forecastability Quality Metric}

To  decide the forecasting question, by comparing alternative versions of the underlying transfer operator kernels for closure of the system, Eq.~(\ref{dev3}) the seemingly obvious way by a Kullback-Leibler divergence  $D_{KL}(q(x-F(y))|| \delta(x-F(y))$ is generally not well defined.  The reason is in part because the $KL$-divergence is not well defined when the support of the second argument is properly contained within the support of the first argument, which will  generally be a problem when stating a $\delta$-function as the second argument and an absolutely continuous function as the first argument.  So in the spirit of transfer entropy, considering $D_{KL}(q(x-F(y))|| \delta(x-F(y))$ may seem relevant but it is not fruitful.

Instead, the Jensen-Shannon divergence gives an alternative that allows several natural associated interpretations.
Let us define the Forecastability Quality Metric, from the $y$-subsystem to the $x$-subsystem,
\begin{equation}\label{FQM}
FQM_{y\rightarrow x}=D_{JS}(q(x-T_x(s))||\delta(x-T_x(s))=\lim_{\epsilon\rightarrow 0} D_{JS}(q(x-T_x(s))||\delta_\epsilon(x-T_x(s)).
\end{equation}
using the notation of Eqs.~(\ref{identa}), and (\ref{form1})-(\ref{form3}).  The influence of $y$ is encoded in the distribution $g$ that has been normalized to the form $q$ from Eq.~(\ref{normalize}).  More will be said on this below.
The Jensen-Shannon divergence is defined as usual, \cite{menendez1997jensen, lin1991divergence, fuglede2004jensen}, 
\begin{equation}
D_{JS}(p_1||p_2)=\frac{D_{KL}(p_1||m)+D_{KL}(p_2||m)}{2},
\end{equation}
where,
\begin{equation}
m=\frac{p_1+p_2}{2},
\end{equation}
the mean distribution.
An important results is that the necessity of support containment is no longer an issue.

The statement of the limit of terms, $\delta_\epsilon$ may be taken as any one of the many variants of smooth functions that progressively (weakly) approximate the action of the delta function, such as for example, 
\begin{equation}
\delta_\epsilon(s)=\frac{e^{-\frac{s^2}{4\epsilon}}}{2\sqrt{\pi \epsilon}},
\end{equation}
but normalized as in Eq.~(\ref{normalize}) for each s related to $x-T_x(s)$.

The Jensen-Shannon divergence has several useful properties and interpretations that are inherited therefore by the FQM.  We summarize some of these here.
$\sqrt{D_{JS}(p_1||p_2)}$ is a metric, stated in the usual sense.  Recall that a function, $d:M\times M\rightarrow {\mathbb R}^+$ is a metric if 1) Non-negative ($d(x,y)\geq 0, \forall x,y\in M$)), 2) Identity and discernible, ($d(x,y)=0$ iff $x=y$), 3) Symmetric, $d(x,y)=d(y,x), \forall x,y\in M$, and 4) Triangle inequality, $d(x,y)\leq d(x,z)+d(z,y), \forall x,y,z \in M$.  The terminology {\it metric} is reserved for those functions $d$ which satisfy 1-4, and {\it distance } while sometimes used interchangeably with metric is sometimes used to denote a function that satisfies perhaps just properties 1-3.  The term {\it divergence} is used to denote a function that may only satisfy property one, but it is only ``distance-like".  So the Kullback-Leibler divergence $D_{KL}$ is clearly not a distance, and only a divergence because it is not symmetric.  

The Jensen-Shannon divergence is not only a divergence but ``essentially" a metric.  More  specifically its square root, $\sqrt{D_{JS}(p_1||p_2)}$ is a metric on a space of distributions, as proved in \cite{endres2003new,  vajda2003new}.  However, nonetheless through Pinsker's inequality there are metric like interpretations of the Kullback-Leibler divergence, that bounds from above, $\sqrt{\frac{D_{KL}(p_1||p_2)}{2}}\geq  \|p_1- p_2\|_{TV}$ by the total variation distance, and for a finite probability space this even relates to the $L^1$ norm, \cite{pinsker1960information, ordentlich2005distribution}.  However a most exciting insight into the meaning of  $1/D_{JS}$ follows the interpretation that relates the number of samples one would have to draw from two probability distributions with confidence that they were selected from $p_1$ or $p_2$ is inversely proportional to the Jensen-Shannon divergence, \cite{tkavcik2008information}.
Thus the Jensen-Shannon divergence is well known as a multi-purpose measure of dissimilarity between probability distributions, and we find it to be particularly useful to build our information flow concept of ``forecasting" as defined, $FQM_{y\rightarrow x}$ by Eq.~(\ref{FQM}) following comparing the operator kernels of Eq.~(\ref{dev3}) as interpreted as conditional probabilities. $FQM_{x\rightarrow y}$ is likewise defined. Finally we remark that the property,
\begin{equation}
0\leq FQM_{y \rightarrow x}\leq 1,
\end{equation}
is inhereted from the similar bound for the underlying Jensen-Shannon divergence.

\section{Example - One Way Coupling and the FQM}\label{examplesec}

Now we specialize the general two oscillator problem Eq.~(\ref{form3}) to specify just one way coupling as an explicit computation of $FQM_{y\rightarrow x}$.
Let $\epsilon_2=0$,
\begin{eqnarray}\label{ident}
x_{n+1}&=&T_x(x_n,y_n)=f_1(x_n)+\epsilon_1 k(x_n,y_n) \nonumber  \\
y_{n+1}&=&T_y(x_n,y_n)=f_2(y_n).
\end{eqnarray}
For simplicity of presentation assume  diffusive coupling,
\begin{equation}
k(x,y)=(y-x),
\end{equation}
 so that,
\begin{equation}\label{diffus2}
x_{n+1}=f_1(x_n)+\epsilon_1 (y_n-x_n)= \tilde{f_1}(x_n)+\epsilon_1 y_n 
\end{equation}
and that, 
\begin{equation}\label{ident7}
\tilde{f_1}(x)=f_1(x)-\epsilon_1 x.
\end{equation} 
Thus we have a special case of a coupled map lattice, \cite{kaneko1991coupled, pethel2006symbolic}.

\begin{figure}[htbp]
\centering
\includegraphics*[width=0.55\textwidth]{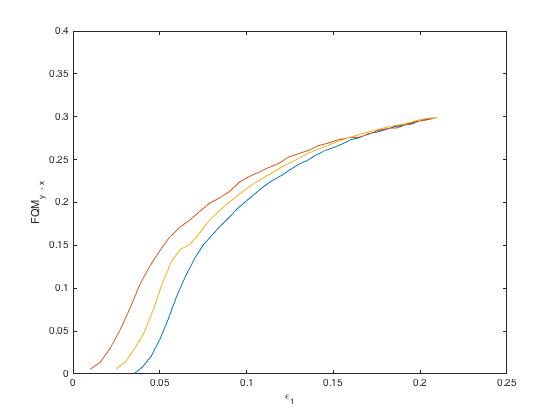}
\caption{Computed $FQM_{y\rightarrow x}$ for coupled logistic maps, in units of bits per time unit of iteration, with varying $\epsilon_1$ increasing as shown on the horizontal axis.  According to the definition of $FQM_{y\rightarrow x}$, Eq.~(\ref{FQM}), Jensen-Shannon divergences are computed for successive approximating values of $\epsilon$ decreasing to zero, $\epsilon=0.035, 0.025, 0.01$ shown.  By definition as a Jensen-Shannon divergence,  Note that $0\leq FQM_{y\rightarrow x} \leq 1$, and $0$ is achieved if the distributions in Fig.~\ref{fig:1} match, which is closely true when $\epsilon=\epsilon_1$ in terms of the coupling.  However, as $\epsilon\rightarrow 0$ for a fixed positive but exceedingly small coupling $0<\epsilon_1<<1$ then the limit is numerically difficult to estimate since the integration becomes singular, and we perform these estimators of the integrals by Monte-Carlo method;  the estimation becomes much more reliable for larger coupling $\epsilon_1>0$ where the direct numerical integration becomes more stable.
}\label{fig:2}
\end{figure}

Further for developing an explicit example, 
\begin{equation}\label{logistic1}
f_1(s)=f_2(s)=4s(1-s),
\end{equation} 
the logistic map.  We take $f_i:{\mathbb R}\rightarrow {\mathbb R}$, but in the uncoupled cases we know that $[0,1]$ is an invariant set for each component.
Since the $y$-subsystem is uncoupled, and we know its absolutely continuous invariant density  in $[0,1]$ is, \cite{lasota1985probabilistic, bollt2013applied},
\begin{equation}\label{logistic2}
\rho(s)=\frac{1}{\pi \sqrt{x (1-x)}}.
\end{equation}
We may take this as the distribution of $y_n\in \Omega_y=[0,1]$ if the $y$-subsystem is taken to be at steady state.
However, we emphasize a {\it steady state distribution need not be assumed} if we assume simply that a distribution of  initial conditions may be chosen from the outside forcing $y$-subsystem.
Since considering the form of the stochastic Frobenius-Perron operator, Eq.~(\ref{twv2}), the outside influence onto the $x$-subsystem looks like the noise  coupling term $\epsilon_1 y_n $ in Eq.~(\ref{diffus2}).  Notice that the distribution of ``noise" $g$ is in fact,
\begin{equation}\label{logistic3}
g(s)=\frac{\rho(\frac{s}{\epsilon_1})}{\epsilon_1},
\end{equation}
which may seem as noise to the $x$-subsystem not knowing the details of a $y$-subsystem, even if the evolution of the full system may even be deterministic.  In fact {\it this may be taken as a story explaining noise generally as the (unknown) accumulated outside influences on a given subsystem.}
So therefore the appearance of ``noise" of $y$-subsystem influence onto $x$ is simply the lack of knowledge of the outside influence onto the not-closed subsystem $x$.  It is a common scenario in chaotic dynamical system that lack of knowledge of states has entropy, and this is the foundation concept of ergodic theory to treat even a deterministic system as a stochastic dynamical system in this sense, as we expanded upon in \cite{bollt2013applied}.  

We see in Fig.~\ref{fig:1} the contrasting versions of Eq.~(\ref{dev2}), $P(x_{n+1}|x_n)= \hspace{-0.08in}? \hspace{0.08in} P(x_{n+1},y_n|x_n)$ associated with contrasting $q(x-F(s))$ to $q_\epsilon(x-F(s))$ corresponding to alternative truths, that the $x-subsystem$ is closed, or open depending on $y$ now considered as a stochastic influence.  The point is: within the transfer operator formalism, the outside influence may be as if stochastic, but nonetheless, the $q$ is a well defined function, and the question of $FQM_{y\rightarrow x}$ is well defined by contrasting the two kernels of the associated transfer operators as if pdfs by the $D_{JS}$ in Eq.~(\ref{FQM}).  

In Fig.~\ref{fig:2} we show a sequence of estimators illustrating $FQM_{y\rightarrow x}$ for Eq.~(\ref{FQM}).  The system shown is relative to the one-way coupled logistic map systems, Eqs.~(\ref{ident})-(\ref{ident7}).  Note that nothing in the current computation requires a steady state hypothesis since considering an ensemble of $y$ values then the resulting integration is well defined by whatever may be the transient distribution.  However as $\epsilon\rightarrow 0$ in the definition, then even though the $FQM_{y\rightarrow x}$ is described by a limit of closed form integrals, they become exceedingly stiff to capture reliable values for both $\epsilon$ {\it and } $\epsilon_1$ small.  In another note, notice that since our discussion in no way requires steady state, the two way coupled problem is just as straight forward as the one way coupled problem, which we highlighted purely for simplicity and pedagogy reasons.  Finally we restate that since $1/D_{JS}$ is descriptive of the number of samples required to distinguish the underlying two distributions, the this sheds lights as  interpretation onto the $FQM_{y\rightarrow x}$ curves in Fig.~\ref{fig:2}  which therefore may be interpreted that as coupling $\epsilon_1$ decreases, the decreasing entropy indicates that significantly more observations, either more time, or more states from many initial conditions, are correspondingly required to decide if there is a second coupling system (open), or the system observed is autonomous (closed).

\section{Postscript and Conclusions}

We have described how noise and coupling of an outside influence onto a subsystem from another subsystem can be formally described as alternative views of the same phenomen. Using these alternative descriptions of this concept, by using the kernels from  deterministic versus stochastic Frobenius-Perron transfer operators to contrast the outside influence of a coupling system as if it were noise, we can explicitly enumerate the degree of information transferred from one subsystem to another.  This is the first time this formalism has been brought to consider information transfer.  We  show furthermore  that motivated by transfer entropy, using the KL-divergence for the transfer operator concept based  in this context produces problems regarding boundedness.  The Jensen-Shannon divergence provides a useful alternative that furthermore comes with several pleasant extra interpretations.

Summarizing outside influence we state the following possibly commuting diagram, pointwise,
\be
\label{eqcommdiagramapp}
\begin{CD}
\rho\in L^1(\Omega) @>P_T>>  \rho'\in L^1(\Omega) \\
@V R_y VV    @VV R_y V  \quad .\\
{\rho_1}\in L^1(\Omega_y) @>P_{T_y} >> {\rho_1}'\in L^1(\Omega_y)
\end{CD}
\ee
where we reiterate that $\Omega=\Omega_x\times \Omega_y$ states the proposed subsystems, and,
\begin{eqnarray}
r_y:\Omega &\rightarrow& \Omega_y, \nonumber \\
(x,y) &\mapsto&y
\end{eqnarray}
denotes a projection function, from the full phase space $\Omega$ to the phase space of the $y$-subsystem, and likewise for the projection $r_x$.
In this formulation the main question of closure, if there is information flow or not, which we have already stated in Eq.~(\ref{dev3}) as 
$q(x-F(y))= \hspace{-0.08in}? \hspace{0.08in} \delta(x-F(y))$, also amounts to asking if advancing the density of states of the full system and then projecting by the operator corresponding to marginalizing, (integrate density onto just $y$ variables, $R_y[\rho(x,y)]=\int_\Omega \rho(x,y dx$ ), is the same as marginalizing first and then advancing by the transfer operator of the subsystem.
\begin{equation}
R_y\circ P_T = \hspace{-0.08in}? \hspace{0.08in}  P_{T_y}\circ R_y.
\end{equation}
In postscript, we already noted that the inverse of the Jensen-Shannon divergence is proportional to the expected number of samples required to distinguish the two distributions.  Therefore, the $FQM_{y\rightarrow x}$ is inversely proportional to the number of samples required to distinguish the degree of coupling influence of the $y$-variables onto the $x$-variables subsystems.  In this sense, in our follow-up work we are planning a practical numerical scheme to associate data observations.  Specifically, the Ulam's method allows for a cell-mapping method to cover the phase space with boxes (or say triangles), and then to collect statistics of transitions, and besides the usual discussion toward invariant density through the eigenvectors of the resulting stochastic matrix, known as Ulam's method, we have already pointed out \cite{bollt2012synchronization} that there is information in this numerical estimate of the transfer operator that can be exploited to compute transfer operator.  However we know realize that the operator  itself bears a great deal of information regarding information flow, and so this points to the idea that FQM might be estimated from data, by using the data to build a stochastic matrix in the spirit of Ulam's method.  That is such a Markov chain model of the process can help distinguish open or closed, but building the transition matrix directly from data, and then applying the FQM, a $D_{JS}$ computation in alternative formulations of the hypothesis.  Therefore we are working toward this for future research, and considering error analysis of the collected statistics has Markov-inequalities (including  Chebyshev inequality) underlying.  Therefore while this more practical data oriented approach is still in the works, what we have offered in this paper is a new view on information flow, that can be understood directly in terms of the underlying transfer operators, and computations of entropies directly therefrom.

\section{Acknowledgements}

The author would like to thank  the Army Research Office (N68164-EG) and the Office of Naval Research (N00014-15-1-2093), and also DARPA.

\section{References}
\bibliography{bibit3}

\end{document}